\documentclass[onecolumn,aps, prl]{revtex4-1}

\usepackage{chngcntr}
\usepackage{geometry}                % See geometry.png to learn the layout options. There are lots.
\usepackage{graphicx}

\usepackage{caption}
\usepackage{subcaption}
\usepackage{float}
\usepackage[T1]{fontenc}
\usepackage{amssymb,amsfonts,amsmath}
\usepackage{stackengine}
\usepackage{xr}

\newcommand{\Fig}[1]{Figure~\ref{#1}}
\newcommand{\fig}[1]{Fig.~\ref{#1}}

\newcommand{\Figs}[1]{Figures~\ref{#1}}

\begin{document}  
\title{Morphology of Proliferating Epithelial Cellular Tissue}

\author{Pranav Madhikar}
\affiliation{Department of Mathematics and Computer Science \&
  Institute for Complex Molecular Systems, Eindhoven University of
  Technology, 5600 MB, Eindhoven, Netherlands }
\author{Jan \AA{}str{\"o}m}
\affiliation{CSC Scientific Computing Ltd, K{\"a}gelstranden 14, 02150 Esbo, Finland}
\author{Bj{\"o}rn Baumeier}
\affiliation{ Department of Mathematics and Computer Science \&
  Institute for Complex Molecular Systems, Eindhoven University of
  Technology, 5600 MB, Eindhoven, Netherlands }
\author{Mikko Karttunen}
\affiliation{Department of Chemistry and Department of Applied
  Mathematics, Western University, 1151 Richmond Street, London,
  Ontario N6A 5B7, Canada}
%\affil[1]{
%Department of Mathematics and Computer Science \&
%  Institute for Complex Molecular Systems, Eindhoven University of
%  Technology, 5600 MB, Eindhoven, Netherlands }
%\affil[2]{CSC Scientific Computing Ltd, K{\"a}gelstranden 14, 02150 Esbo, Finland}
%\affil[3]{Department of Chemistry and Department of Applied
%  Mathematics, Western University, 1151 Richmond Street, London,
%  Ontario N6A 5B7, Canada}

\date{November 16, 2018}
% Activate to display a given date or no daate

%\counterwithin{figure}{section}
%\bibliographystyle{nature}

\begin{abstract}
We investigate morphologies of proliferating cellular tissue using a
  newly developed numerical simulation model for mechanical cell
  division. The model reproduces structures of simple multi-cellular
  organisms via simple rules for selective division and division plane
  orientation. The model is applied to a bimodal mixture of stiff
  cells with a low growth potential and soft cells with a high growth
  potential.  In an even mixture, the soft cells develop into a tissue
  matrix and the stiff cells into a dendrite-like network
  structure. For soft cell inclusion in a stiff cellular matrix, the
  soft cells develop to a fast growing tumour like structure that
  gradually evacuates the stiff cell matrix. With increasing inter-cell
  friction, the tumour growth slows down and parts of it is driven to
                                    self-inflicted cell death.
\end{abstract}

\maketitle

%%%%%%%%%%%%%%%%%%%%5

Morphology and dynamics of proliferating cells are among the
fundamental issues at different stages of cellular
development~\cite{Paluch2009,Li2012,Heisenberg2013,Ragkousi2014,Gibson2014,Sanchez-Gutierrez2015}.
They are controlled by a number of factors, but from the physical point of view,
morphology is tightly coupled to inter-cellular force transmission, see e.g.,
Refs.~\cite{Rauzi2008,Tambe2011,Sanchez-Gutierrez2015};
mechanical forces have been shown to be important in cancer development and it has been
suggested that tumour growth may even be arrested by inter-cellular mechanical forces~\cite{Helmlinger1997,Kumar2009}.
Among the many complications in investigating force transmission are that at
their embryonic state, cells may not yet have developed junctions and
may display more fluid-like behaviour, and that cell-cell adhesion
depends on the cell types~\cite{Foty1996,Lecuit2005,
  Rauzi2008, Guillot2013}. Junctions are crucial in
cell-to-cell stress transmission~\cite{Rauzi2008, Liu2010, Tambe2011,Pontani2012} but it is, however,
challenging to probe the individual junctions experimentally.

From a coarse-grained point of view, that is, ignoring chemical details and treating
cells as elastic objects, cellular systems can be seen as
disperse grand canonical soft colloidal systems under evolving pressure
applied non-uniformly throughout the system.
Several studies have tried to capture aspects
of growing soft colloidal systems (e.g. for modelling tissue growth) at different
levels~\cite{Astrom2006,Jones2012,Gonzalez-Rodriguez2012, Kursawe2018} but even in simple
systems many fundamental questions remain open including the precise nature
of colloidal phase diagrams when colloids are soft with size
dispersity~\cite{Fernandez2007}, and structure selection via
self-assembly~\cite{Frenkel2011}. Cellular systems are more
complex since they exhibit additional behaviours such as cell
growth and division, they have varying mechanical properties (e.g. elasticity
and cell-cell adhesion) and their response to external stimulus may be sensitive to the
local environment.

Dimensionality has an important role in particular in regulation of intra- and inter-cellular forces at
different levels, see e.g.
Refs.~\cite{Charnley2013,Hernandez-Hernandez2014,Lesman2014,Osterfield2017,Owen2017}.
Some systems, such as epithelial tissues and \textit{Drosophila} wing discs,
are inherently two dimensional which gives them distinct morphological
properties due to the nature of cellular packing, and transmission of and response to
forces~\cite{Gibson2009,Guillot2013,Gibson2014}.
In addition, jamming can be very strong in two dimensional systems and
varying stiffness and inter-membrane friction is a step towards
investigating jamming in cellular systems~\cite{Sadati2013,Reichhardt2014}. Our
main focus is on the above effects in systems consisting of hard cells in a soft matrix and vice
versa. Besides being important in understanding the mechanisms of cell movement
under pressure~\cite{Tambe2011}, such situations have been proposed to be important in tumour
growth~\cite{Helmlinger1997,Kumar2009} -- cancer cells are often softer than healthy
cells~\cite{Palmieri2015,Alibert2017} although the opposite has also been reported~\cite{Suresh2007,Nguyen2016a}. Cell stiffness, its measurements and connection to cancer metastasis have been recently reviewed by Luo et al.~\cite{Luo2016}.

One of the intriguing questions in cell division is: Why do cells exhibit diverse morphologies upon division and growth? In addition to uniform structures, a plethora of structures with various mechanisms and division modes have  been suggested but the issue remains largely
unresolved~\cite{Pilot2005,Gibson2006,Guillot2013,Herrero2016}.
To illustrate how different morphologies can arise, consider
cyanobacteria.  Figures \ref{fig:exp_cyano}a)-c) show \textit{Anabanea
  circinalis} and \fig{fig:exp_cyano}d) \textit{A. flos-aquae}.
Both of them can be approximated  as quasi one-dimensional structures.
In contrast to most other cells, however, cyanobacteria have  continuous outer membranes shared by the whole filament %(\Fig{fig:exp_cyano}a)-d)) 
consisting of multiple cells~\cite{Herrero2016,Zhang2018}. The inner membranes, however, belong to individual bacteria only. It has been suggested that this together with specialized junctions leads to filamentous structures (there are also subtleties related to, e.g., size selection of the filaments)~\cite{Herrero2016,Zhang2018}.

In our previous study~\cite{Mkrtchyan2014}, the orientation of the division line was selected randomly.
The simplest way to model cyanobacteria morphologies is to choose the orientation of the division plane in such a way that it is approximately parallel between neighbouring cells. Although an approximation, this approach should be able to produce similar morphologies.

Typically, two dimensional systems exhibiting uniform non-directional growth have been studied
using vertex models, see, e.g. Refs.~\cite{Graner1992, Staple2010,
  Fletcher2010} and Refs.~\cite{Merks2005, Jones2012} for reviews. One dimensional systems, however, have gained lesser attention and the models are typically of reaction-diffusion type with fixed geometry and size as discussed extensively in the review by Herrero et al.~\cite{Herrero2016}

We employ the two-dimensional CeDEM (the Cellular Discrete Element Model) to investigate
tissue morphologies in one and two dimensions. Full  details and derivation of the model
are provided in Ref.~\cite{Mkrtchyan2014} but to summarize, in CeDEM
the cell membrane is discretized as beads connected by bonds of
stiffness $K^\mathrm{spr}_i$.

Cellular growth is controlled by a growth
pressure and division by a threshold in
cell area (above which cells divide) and the orientation of the
cell division line.
Importantly, CeDEM  allows the topology (the polygonal distribution) to vary spontaneously~\cite{Mkrtchyan2014}.

Here, we extend CeDEM for simulations of different cell types using three simple approaches:
\begin{enumerate}
\item changing the cell division line orientation,
\item changing cell stiffness, and, finally,
\item  changing the friction between cell membranes; in CeDEM cell membrane and cytoskeleton are treated as a coarse-grained single object.
\end{enumerate}
%%%%

In our previous study, the orientation of the division line of each cell was chosen randomly and it was constrained to pass through the centre of mass of the cell undergoing division. This results in tissue growth such that it fills
the available space roughly uniformly~\cite{Mkrtchyan2014}.
Modification 1) above allows for simulations of cyanobacteria-like structures shown in~\fig{fig:exp_cyano}.

Modification 2) allows for simulations of
different types of cells. 
As mentioned above, cancer cells typically considered to be softer than the matrix cells. Softness, or higher malleability, is typically associated with the invasiveness of cancer cells~\cite{Luo2016}. This has recently been challenged by 
Nguyen~\cite{Nguyen2016a} et al. who measured Young's modules of pancreatic cancer cells using different cell lines and found the stiffer (than the matrix cells) cells to be more invasive than the softer cancer cells. Whether this is purely mechanical or due to simultaneously occurring biological processes remains unclear.
Here, we use two types:  1) \textit{Type1}, stiff cells with a low growth
potential with stiffness $K^\mathrm{spr}_1$. The low growth potential means
that the cell membrane is so stiff that the applied pressure is barely
enough to grow the cell to a size above the division
threshold. Therefore, if the cell is even lightly squeezed between
other cells it will not divide before force equilibrium is reached and
growth stops. 2) \textit{Type2}, soft cells with a high growth potential with
stiffness $K^\mathrm{spr}_2$. These cells have a high growth potential which
means that cell membrane stiffness is so low that the cell area easily
grows beyond the division size. The cells are identical in all other
ways except their stiffness.

Finally, modification 3) allows for comparisons of systems of cells with different inter-membrane
friction coefficients. 
Cell-cell friction and its importance in mechanotransduction has recently been reviewed by Angelini et al.~\cite{Angelini2012}.
Inter-membrane friction is
modelled as%
\[
  \vec{F}^\mathrm{ext}_i = -\mu\vec{v}_{ij},
\]%
where $\mu$ is the friction coefficient and $\vec{v}_{ij}$ is
the component of the relative velocity between two membranes
tangential to the cell that bead $i$ belongs to.
We compare systems where $\mu=0.0$, that is,  cells
do not interact very much with their neighbours, and strongly interacting cells with $\mu=20.0$.

\section*{Results}

\subsection*{Quasi-one-dimensional morphology}
\begin{figure}[ht]
	\centering
	\includegraphics[width=\columnwidth]{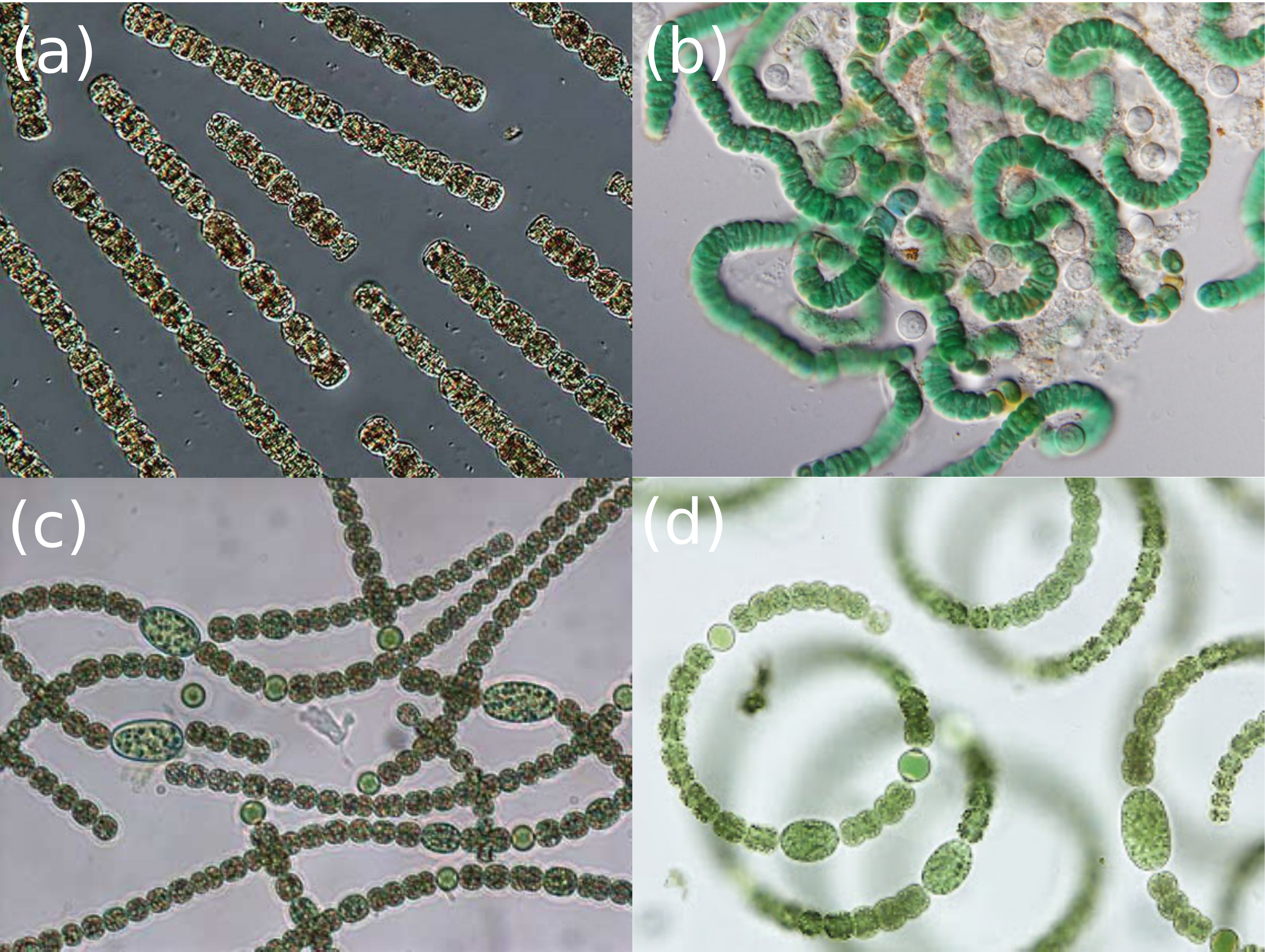}
	\caption{ a) \textit{A. planctonica}. Image with
		permission from the Laboratory of Phytoplankton
		Ecology~\cite{Znachor2004}. b) \textit{A. laxa}. Image
		with permission from the A. Braun Culture Collection of
		Autotrophic Organisms~\cite{Josef1991}. c) \textit{A.
			circinalis}. Image with permission from the Kudela Lab,
		University of California Santa Cruz~\cite{Kudela2018}. d)
		\textit{A. flos-aquae}. Image with permission from
		Demarteau/Aquon~\cite{Demarteau2011}.}
	\label{fig:exp_cyano}
\end{figure}

\begin{figure}[ht]
  \centering
  \includegraphics[width=\columnwidth]{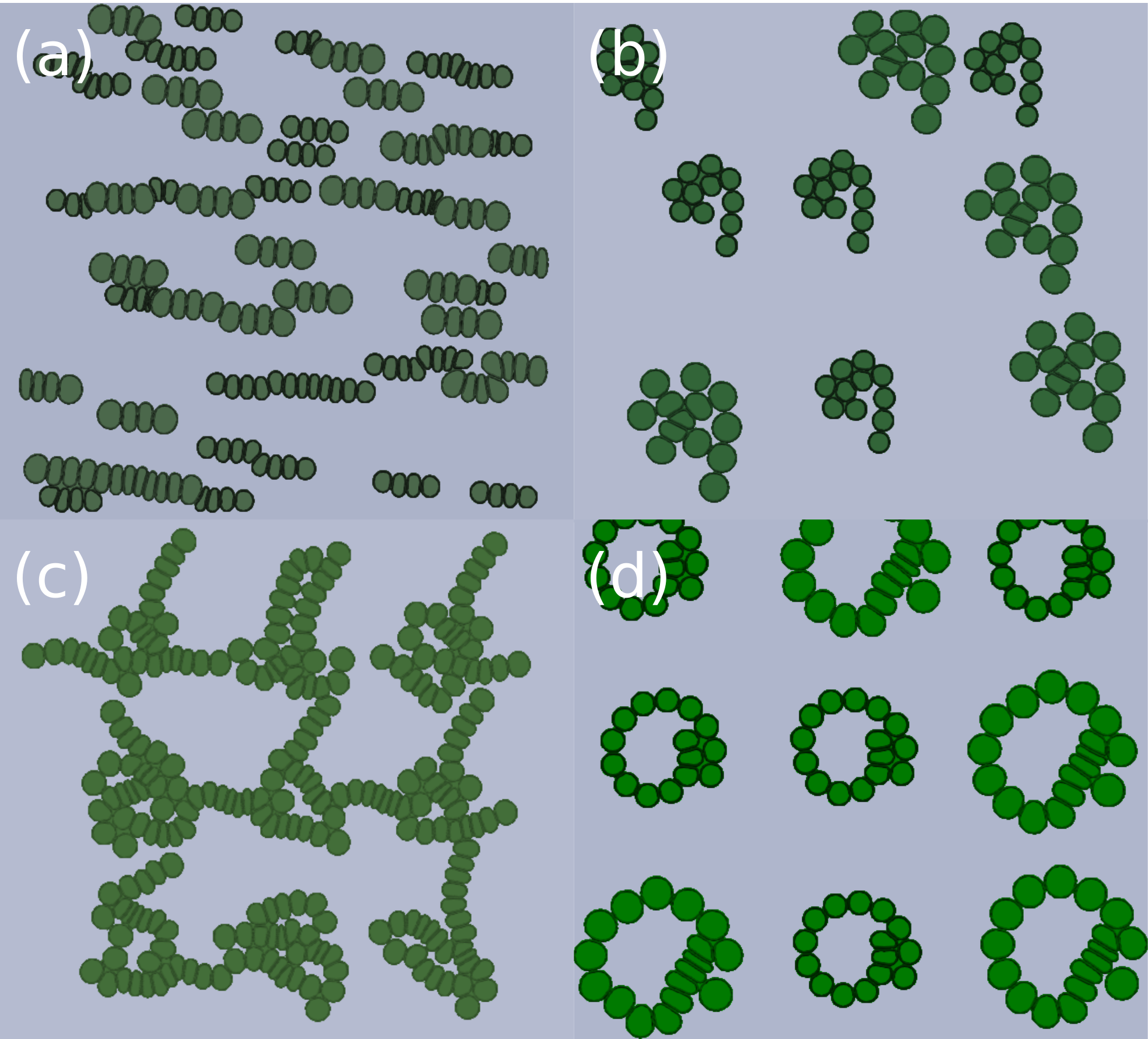}
  \caption{The type of cell can be changed by changing
    the division plane choosing rules. a) Division planes parallel at
    each generation, b) division plane turning left at an increasing rate,
    c) small clusters of original cells dividing at perpendicular and
    constant orientation angles, and d) division plane turning left at
    constant rate.}
\label{fig:cyano}
\end{figure}

We start from (quasi-) one-dimensional systems and
compare the structures from experimental systems (Fig.~\ref{fig:exp_cyano}) and simulations (Fig.~\ref{fig:cyano}). First,
instead of just dividing all cells that are above some threshold area,
we allow a single cell to divide only once.
Thus, only the youngest cells are allowed to divide similar to budding growth in
bacteria~\cite{Wang2017}. Additionally, we make the division plane
non-random. Different scenarios lead to morphologies as shown in Fig.~\ref{fig:cyano}:
Keeping the division plane parallel for each generation leads to morphologies similar to \textit{A. planctonica} (compare Figs.~\ref{fig:exp_cyano}a and \ref{fig:cyano}a).
Letting the division line rotate slightly more by every generation produces~\fig{fig:cyano}b. Allowing all cells to divide and letting them divide along two perpendicular lines produces~\fig{fig:cyano}c, approximating the morphology of \textit{A. laxa} in Fig.~\ref{fig:exp_cyano}b.
Finally, constant rotation every generation leads to \fig{fig:cyano}d, which is structurally similar to Figs.~\ref{fig:exp_cyano}b and \ref{fig:exp_cyano}d.

Although the morphologies in Fig.~\ref{fig:cyano} are created by the simple rules as discussed above, and there may well be other rules that lead to similar morphologies, it is important to keep in mind that to arrive to such structures real systems have molecular mechanisms that lead to the emergence of such structures. The microscopic molecular level mechanisms are effectively manifested as rules at the macroscopic level. The exact mechanisms as why filamentous shapes form remain to be resolved, but current evidence shows that septal junctions have an important role~\cite{Herrero2016,Flores2016}.

\subsection*{Soft and stiff cells in 2D}

\begin{figure*}[ht]
	\centering
	\includegraphics[width=\textwidth]{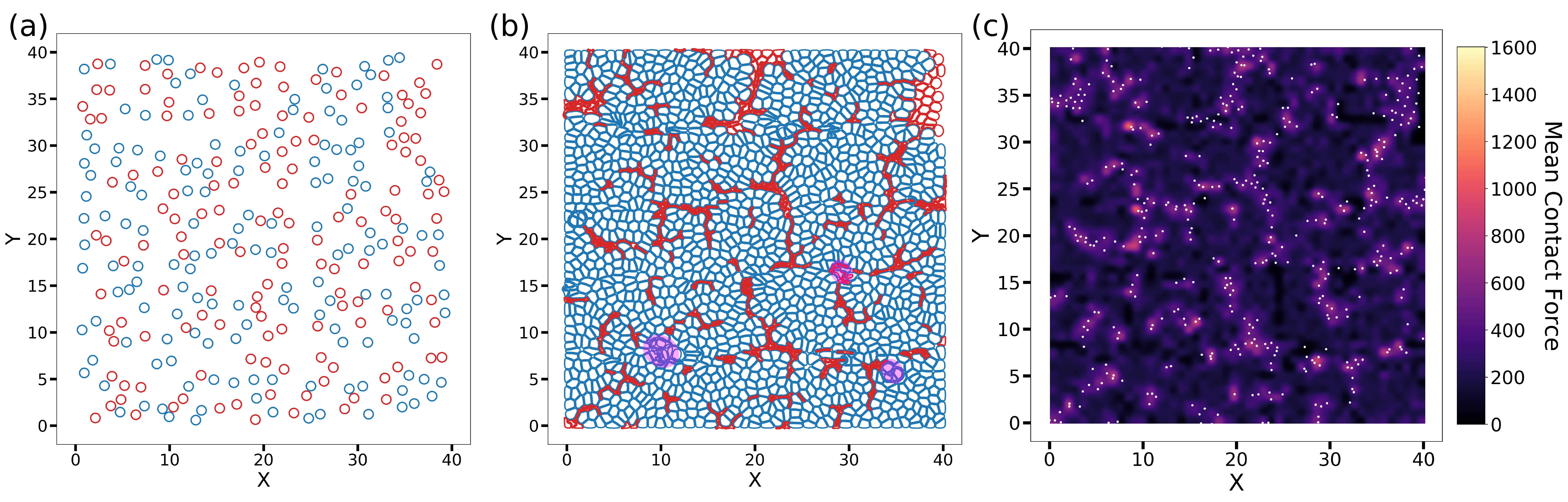}
	\caption{a) Initial configuration.
          Stiff cells are depicted in red, and soft in blue. Growth
		is simulated from this state until confluence.  b) A confluent
		tissue soft and stiff cells. Stiff cells form dendrite or
		vein-like structures in a matrix of soft cells. The regions marked
		with light purple are areas where cells interpenetrate and cell
		death could occur --- though death is not simulated by CeDEM. c)
		Contact force distribution in the same tissue. Large contact
		forces are located at stiff cells and at boundaries between soft
		and stiff cells. White markers are the centres of masses of the
		stiff cells.}
	\label{fig:force_distrib}
\end{figure*}

We now focus on two-dimensional larger and denser samples of cells
with two cell types, stiff (\textit{Type1}) and soft (\textit{Type2}),
in the same system. We assume that softer cells are tumour cells. This
assumption is based on the fact that cancer cells tend to be
softer~\cite{Palmieri2015,Alibert2017}. The initial setups for simulations of such systems were created with
equal proportions of \textit{Type1} (red) and \textit{Type2} (blue)
cells, see \Fig{fig:force_distrib}a). Growth is simulated with
identical parameters for all cells, except membrane stiffness, until
confluence. \Fig{fig:force_distrib}b) shows the tissue structure at
the end of the simulation with the system mostly filled
with soft cells (blue) while the stiff cells (red) are
compressed into dendrite-like structures. Another distinct feature is that the cells
interpenetrate in the regions  marked with
light purple in \fig{fig:force_distrib}b). 
This type of behaviour occurs in diverse systems 
as shown by Eisenhoffer \textit{et al.} for canine, human and zebrafish
epithelial cells~\cite{Eisenhoffer2012} and
discussed at length in the review by Guillot and
Lecuit~\cite{Guillot2013} (see in particular Fig.~2 in Ref.~\cite{Guillot2013}). 
The forces can become so
high that the cell membranes practically intersect each other. These
cells would be good candidates for cell death. Experiments
have also suggested that for live cells, such conditions may lead to
pathologies~\cite{Eisenhoffer2012}. 
CeDEM does not currently support cell
death in terms of cells disintegrating and disappearing from the
system. 
Cells do, however, get squeezed into very small space and
division ceases in the purple regions of Fig.~\ref{fig:force_distrib}b).

\Fig{fig:force_distrib}c) shows a smoothed histogram of the average
inter-membrane (or contact) forces between cells.  The white dots show
the centres of masses of the stiff cells. The peaks in the contact force
distribution correlate highly with the locations of the stiff
cells indicating that \textit{Type2} cells overwhelm \textit{Type1} cells as the
tissue grows and also that the system imposes higher stresses on the stiff collapsed cells.

\begin{figure*}[ht]
	\centering
	\includegraphics[width=\textwidth]{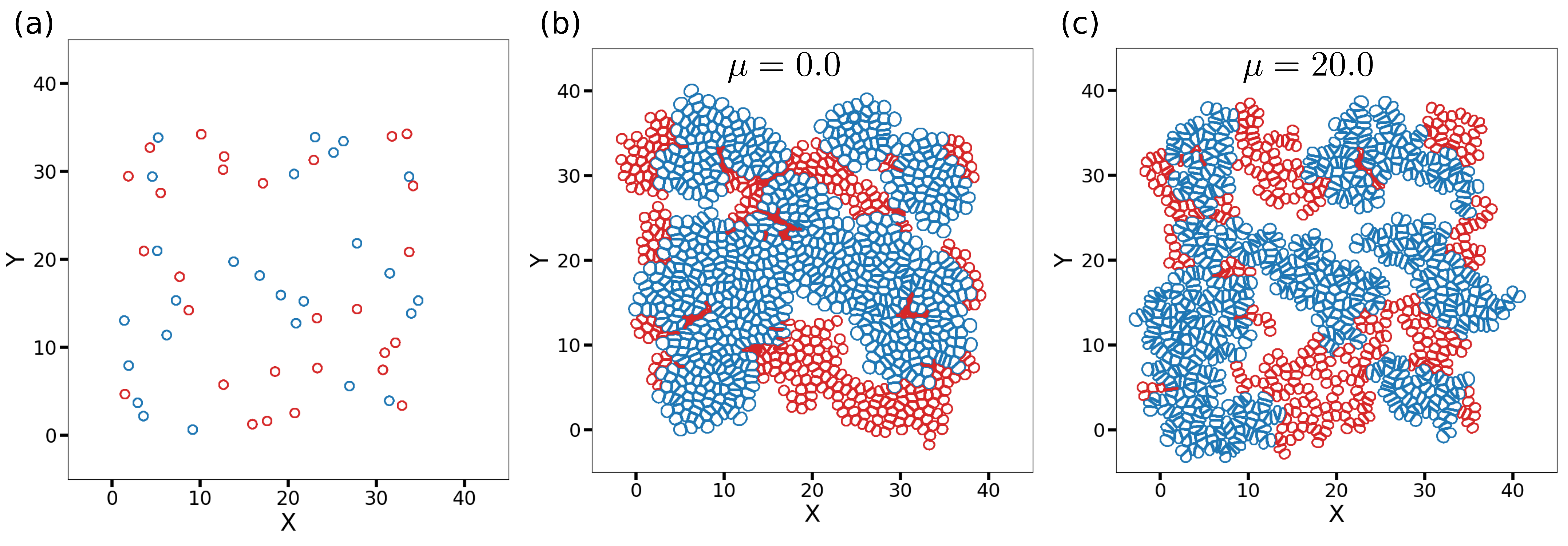}
	\caption{Morphologies of simulated cells with different
		inter-cellular friction. a) Initial conditions stiff cells are in
		red and soft cells in blue. The two are in equal proportions. b)
		Morphology of zero inter-cellular friction cells. c) Tissue with
		high-friction cells ($\mu=20.0$). Both cases were simulated for a
		time corresponding to 10 division cycles.}
	\label{fig:sim_hl_mu}
\end{figure*}

At this point, we ask the question if this collapse of stiff cells can
be mitigated by making their interactions stronger. This can be examined by
changing the magnitude of inter-membrane friction $\mu$. Since cells
need to find space to grow, they need to slide past each other into
empty regions. In other words, higher friction induces
jamming between the cells which means that they easily get squeezed
between each other and therefore reaching the division threshold area
takes a much longer time. The softer cells will also need to
counteract this effect to grow. Figure~\ref{fig:sim_hl_mu}
shows a similar simulation setup as before,
except with different values of $\mu$. Figure~\ref{fig:sim_hl_mu}a) shows the
initial conditions, and Figs.~\ref{fig:sim_hl_mu}b) and c) show the final state
at $\mu=0.0$ and $\mu=20.0$, respectively. 

At low inter-membrane
friction ($\mu=0.0$), there are more cells at the end of the
simulation indicating that growth is faster. The high inter-membrane friction
system ($\mu=20.0$) is more porous with slower growth.
The friction-less system (\fig{fig:sim_hl_mu}a)) corresponds to very
early stages of development when junctions have not yet developed.
The latter system corresponds (\fig{fig:sim_hl_mu}b)) to when cell adhesion molecules have
developed. In both cases, the simulations were run for interval of time corresponding to 10 division cycles.

\begin{figure}[ht]
	\centering
	\includegraphics[width=\columnwidth]{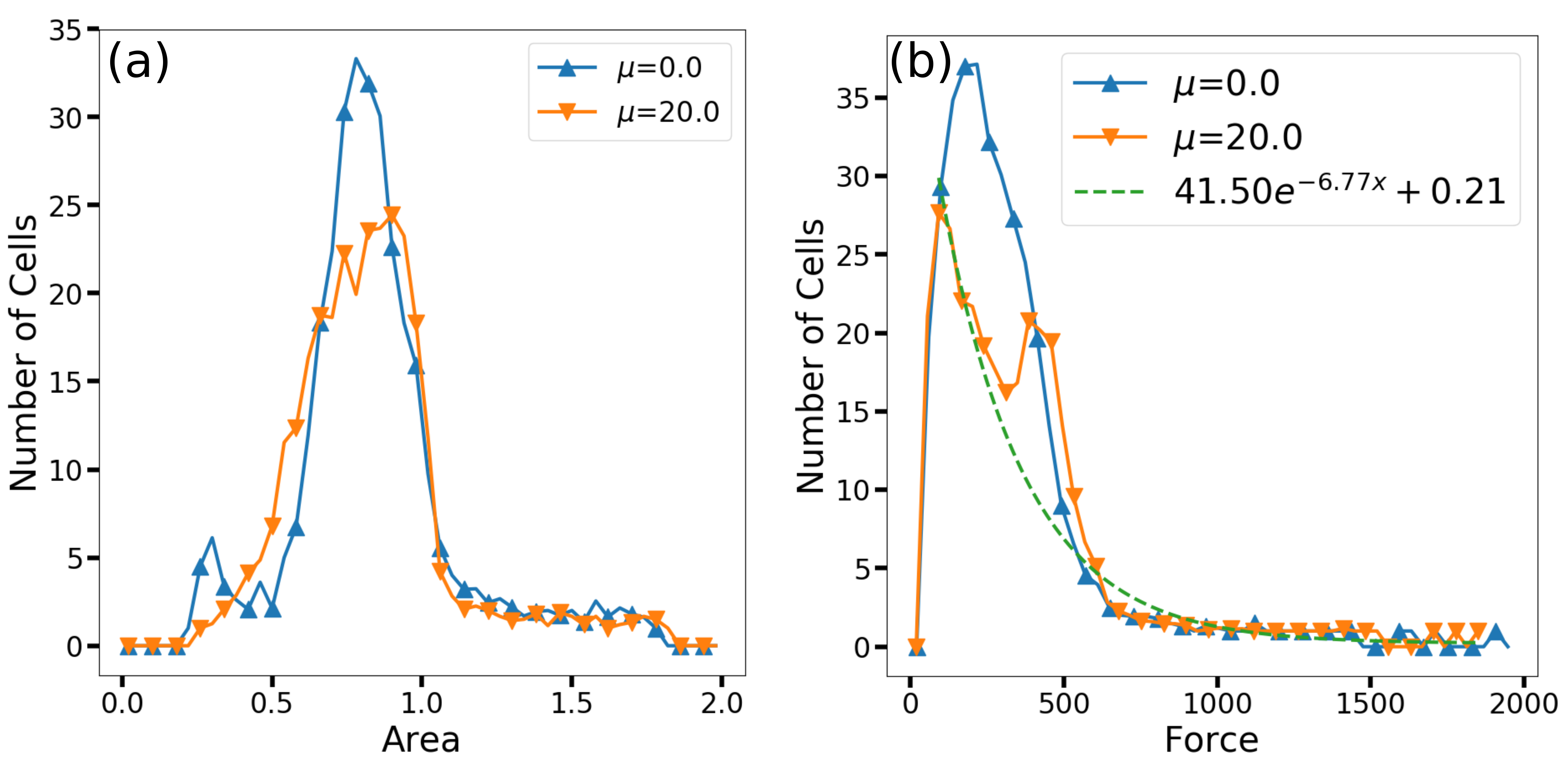}
	\caption{a) Number distributions of cell areas. There is a peak at
		low areas corresponding to collapsed stiff cells. b) Number
		distribution of inter-cell forces at high friction ($\mu=20.0$) and low
		friction ($\mu=0.0$). The green dashed line is an exponential fit to
		the $\mu=20.0$ case, ignoring the second peak.}
	\label{fig:area_force_hist}
\end{figure}

To investigate further, we study the sizes of the cells in each case
and the forces that are acting on the cells. \Fig{fig:area_force_hist}
shows the number distributions of cell area
(\fig{fig:area_force_hist}a)) and the total force
(attractive, repulsive, and friction) that each cell feels due to its
neighbours (\fig{fig:area_force_hist}a)). Both distributions display
lower total number of cells in the high friction tissue. The peak
in area distributions is just below 1.0, which is due to the threshold
division area ($A^{div} = 1.0$). Some of the cell areas have grown
past this limit as cell division occur only at discrete time intervals
in CeDEM so some cells are larger. The $\mu=0$ distribution shows a
small peak at $A\approx 0.2$, which is due to the higher number of
collapsed cells in the low friction system. The large-area peak
represents the soft-cell majority, and its shape is approximately
Gaussian, consistent with the observation from simulations of non-dividing soft
colloids~\cite{Astrom2006}. For $\mu=20.0$ the
distribution has not yet developed two peaks and there are some cells
that can grow rapidly in the sparse areas of the packing.

\begin{figure}[ht]
	\centering
	\includegraphics[width=.65\columnwidth]{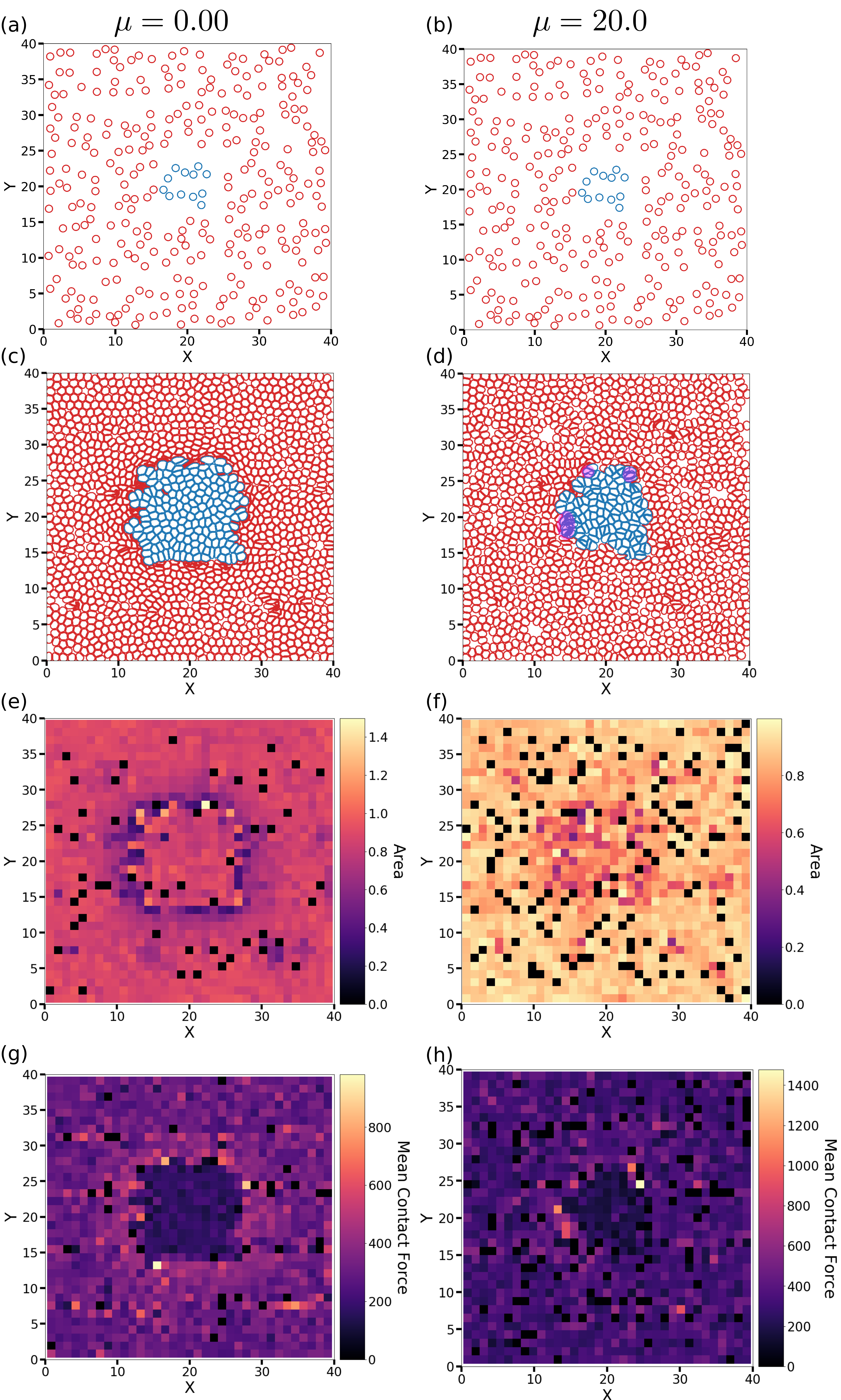}
	\caption{a) Inclusion of \textit{Type2} (blue) cells in a matrix of
		\textit{Type1} (red) cells, $\mu=0.0$, b) $\mu=20.0$. c) ,d)
		configurations for the two cases at confluence. e) Spatial
		cell size distribution, $\mu=0.0$, f) $\mu=20.0$. g) Spatial
		contact forces distribution $\mu=0.0$, h) $\mu=20.0$. Black
		squares: voids in e)--h).}
	\label{fig:snap_area_force}
      \end{figure}

Finally, we study the case of a cluster of soft cells
inside a matrix of stiff cells, \Fig{fig:snap_area_force}.  We
first investigate the case when $\mu=0.0$. In this case, \textit{Type2} cells
with the larger growth potential continue to proliferate even when the
tissue approaches the state of being space-filling, while \textit{Type1}
proliferation almost stops. This leads to a tumour-like growth of
\textit{Type2} cells and compression of \textit{Type1} cells at the tumour
boundary.

\begin{figure}[ht]
	\centering
	\includegraphics[width=\columnwidth]{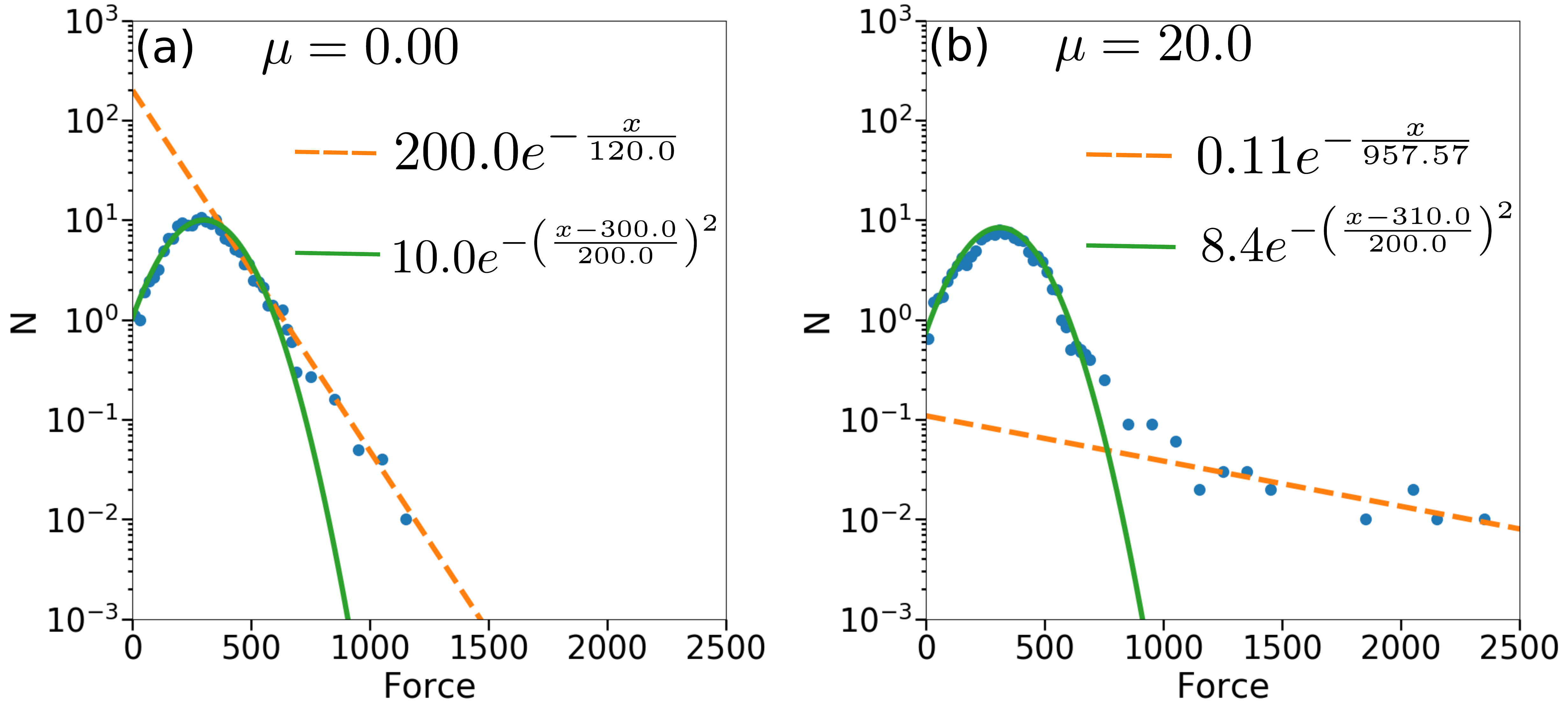}
	\caption{Distributions of inter-cellular forces when a soft
		\textit{Type2} cell is introduced into a tissue consisting of
		stiff \textit{Type1} cells. a) $\mu=0.0$ and b) $\mu=20.0$. The
		distribution of both is similar except the $\mu=20.0$ case is
		slightly wider at forces between 0-1000 and there is a smaller
		number of cells in b) that experience high forces. See the text
              for a detailed discussion and relation to jamming.}
	\label{fig:mixed_force_distrib}
\end{figure}

\Figs{fig:snap_area_force}a) and b) show the initial configurations of
this type of simulation for $\mu=0.0$ and
$\mu=20.0$. Figures~\ref{fig:snap_area_force}c) and d) show the
morphologies for each case at confluence. The faster growth of the
tumour at $\mu=0.0$ is clearly visible. \Figs{fig:snap_area_force}e) and
f) show spatial size distributions in the two cases. The dark
regions in the histograms correspond to pores in the system. Cell
sizes are roughly equal within the tumour and inside the matrix. Along
the tumour boundary, however, the matrix cells are compressed and the
tumour cells are enlarged. This effect is seen in both cases but is
much more pronounced when $\mu=0.0$. Lastly,
\Figs{fig:snap_area_force}g) and h) show the distribution of the mean
contact forces. Inside the tumour the contact forces are low, and the
largest forces are seen scattered on the tumour boundary. Again this
effect is more pronounced in the $\mu=0.0$ case. 
However, even though
there are fewer stressed cells at higher $\mu$, the few cells that are
stressed feel higher stresses. This is quantified in
\Figs{fig:mixed_force_distrib}a) and b) which show the population
distribution over contact force. 
in both of the current cases we see an exponential
tail representing the small population of stressed cells at the
tumour boundary. There are more cells that feel higher stress
in \fig{fig:mixed_force_distrib}a) with $\mu=0.0$, than in
\fig{fig:mixed_force_distrib}b) with $\mu=20$. However, these cells feel
more stress at higher inter-membrane friction.

%%%%
There is only very limited amount of data available for force distributions
in proliferating systems. However, they have been measured for soft colloidal systems
under compression. It is well established that the distribution has an exponential
tail in the vicinity of the jamming transition~\cite{Erikson:02eb,Astrom2006,Jose2016}. It has also been recently shown experimentally by 
Jose et al. by using 3-dimensional packings of soft colloids that
the distribution well above the jamming distribution becomes Gaussian~\cite{Jose2016}.
As Fig.~\ref{fig:mixed_force_distrib} shows, the exponential tail is present in
our two-component 2-dimensional cell systems both at zero and high friction. The fact that the cells
grow also means that their volumes are not conserved (in contrast to experiments with
typical colloids). This is also the case for the cells that are being pushed and compressed by
their neighbours as is evident from the snapshots in Figs.~\ref{fig:sim_hl_mu} and \ref{fig:snap_area_force}. What is clearly different here is the distribution at
low forces: the exponential is preceded by a Gaussian distribution. Gaussian
peak has been observed in simulations of soft colloids in two dimensions with zero friction~\cite{Astrom2006}.
In contrast, in the three dimensional experiments of Jose et al. the low force part of
the distribution remained almost flat except well above jamming transition.
This may have to do with the hardness of the particles: Erikson et al. studied
materials of different hardness and the force distribution at lower forces depends
strongly on hardness~\cite{Erikson:02eb}. In addition,
van Eerd et al. have reported faster than exponential decay from their high
accuracy Monte Carlo simulations~\cite{Eerd2007} although the deviations can
be very hard to detect without high accuracy sampling methods.

Here,
peaks in the
distribution that develop at relatively large forces within the bodies
of stiff cells. This becomes particularly evident when a tissue of
soft and stiff cells becomes so dense that it approaches
space-filling. In this case, almost all stiff cells collapse and form
narrow veins or dendrites. With a suitable initial mixture of stiff
and soft cells, the soft cells form a matrix with a percolating
fractal network of stiff cells which covers only a small fraction of the total area but penetrates almost everywhere(\fig{fig:force_distrib}b)).  This shows a possible pathway
for the formation of signalling and transport networks in a simple
multi-cellular system.

The results in \Fig{fig:snap_area_force} show that the softer cells
introduced into matrices of stiffer cells grow faster when
inter-membrane friction is low; weaker cell-cell interactions provide
conditions for easier growth. This also suggests that inter-cellular
interactions can be an indicator of how well epithelial tissue can
diminish the growth of rogue cells that have a higher growth potential.

\section*{Conclusions}

In this work we use the CeDEM model to study how filamentous growing bacteria can
create varied quasi-one-dimensional morphologies. We show that
modulating the cell division line orientation can be one of the ways
such morphologies can arise. What determines the division line
orientation is, however, an open question but cell-cell junctions have been indicated having an important role~\cite{Herrero2016}. We propose considering
simple division line placement rules as a
possible effective manifestation of  yet unknown microscopic mechanisms.

We then studied larger, denser systems of cells of two different
types in two dimensions. Cell populations are differentiated by their membrane/cortex stiffness. We showed that this simple difference is enough, provided internal pressure is
identical for both, to favour soft cell growth. Even if a few soft
cells are introduced into a matrix of stiff cells, it is enough for the
softer cell to grow rapidly. This effect can be mitigated by a higher
interaction strength between cells.
 Both of the effects above required some modifications of the CeDEM
 model presented before in Refs.~\cite{Astrom2006, Mkrtchyan2014}.
We also studied the force distributions which shows similarities to non-proliferating
soft colloidal systems. Although not studied here in detail, the model allows for
tuning the cell-cell friction, an issue that recently been raised by Vinuth and Sastry
for shear jamming~\cite{Vinutha2016}.

The existing paradigm for the softness of cancer cells has been challenged by  Rowat and coworkers who have shown that stiff cancer cells can be more invasive than soft ones~\cite{Nguyen2016a}. They have also shown that cells experience significant strain hardening. The precise role of it remains to be resolved~\cite{Nyberg2017}. Models such as the current one may be helpful in isolating and identifying the purely mechanical processes and their importance for a collection of cells and related them to other soft matter systems.

\section*{Acknowledgements}
MK would like to thank the Discovery and Canada Research Chairs Programs of the Natural Sciences and Engineering Research Council of Canada (NSERC) for financial support.

%%%END OF MAIN TEXT%%%

%%%REFERENCES%%%
%\bibliography{references} %You need to replace "rsc" on this line with the name of your .bib file
%\bibliographystyle{rsc} %the RSC's .bst file

\providecommand*{\mcitethebibliography}{\thebibliography}
\csname @ifundefined\endcsname{endmcitethebibliography}
{\let\endmcitethebibliography\endthebibliography}{}

\newpage

\end{document}